\documentclass[12pt]{article}
\usepackage{graphicx}
\usepackage{amssymb}
\usepackage{amsfonts}
\usepackage{amsmath}

\begin{document}

%\issuearea{Physics of Elementary Particles и Atomic Nuclei. Theory}

\title{\textbf{Superpotential method for $F(R)$ cosmological models}}

\author{S.Yu.~Vernov$^{a,}$\footnote{E-mail: svernov@theory.sinp.msu.ru}, V.R.~Ivanov$^{b,}$\footnote{E-mail: vsvd.ivanov@gmail.com},  E.O.~Pozdeeva$^{a,}$\footnote{E-mail: pozdeeva@www-hep.sinp.msu.ru} \\
\small $^a$ Skobeltsyn Institute of Nuclear Physics, Lomonosov Moscow State University,\\ \small  Leninskie Gory~1, 119991, Moscow, Russia\\
\small $^b$ Physics Department, Lomonosov Moscow State University,\\
\small Leninskiye Gory~1, 119991, Moscow, Russia }
\date{  \  }
\maketitle

\begin{abstract}
     We construct the $F(R)$ gravity models with exact particular solutions using the conformal transformation and the superpotential method for the corresponding models in the Einstein frame. The functions $F(R)$ are obtained explicitly. We consider exact solutions for the obtained $R^2$ gravity model with the cosmological constant in detail.
  \end{abstract}
\section{Introduction}

The $F(R)$ gravity is one of the most popular generalizations of the general relativity~\cite{Sotiriou:2008rp,Tsujikawa:2010,Faulkner:2006ub}. The $F(R)$ gravity models are actively used to describe different epochs of the Universe evolution. For example, the Starobinsky $R^2$ inflationary model~\cite{Starobinsky} (see also~\cite{Mijic:1986iv}) leads to the predictions that do not contradict to the observation data~\cite{Planck2018}.
Dark energy  $F(R)$ models are actively investigated and analyzed, for example, in papers~\cite{Capozziello:2003gx,Dolgov:2003px,Hu:2007nk,Starobinsky:2007hu,Tsujikawa:2007xu,Bamba:2008hq,Ali:2010zx,Capozziello:2018wul,Arbuzova:2019ptx}. Moreover, there are $F(R)$ gravity models that can describe both inflation and the late time cosmic acceleration~\cite{Nojiri:2003ft,Motohashi:2010tb,OdintsovRev}.

Exact solutions play an important role in cosmology and the search of integrable $F(R)$ models as well as models with exact particular solutions is an interesting problem~\cite{Paliathanasis:2017apr,Muller:2017nxg}.
A $F(R)$ gravity model can be transformed into a model with a minimally coupled scalar field with a canonical kinetic term by the metric and scalar field transformations~\cite{Maeda:1988ab}. There are a few methods to construct models with minimally coupled scalar fields with exact cosmological solutions. One of the popular methods is the superpotential one~\cite{SalopekBond,Muslimov} (also known as the Hamilton--Jacobi method or the first-order formalism). In this paper, we generalize this method on $F(R)$ gravity models to get such models with exact solutions. We show a few examples of such models with an explicit dependence of $F(R)$.

\section{The corresponding Einstein frame models}

Let us consider an $F(R)$ gravity model:
\begin{equation}
\label{actionFR}
    S_{R} =\int d^4 \tilde{x} \sqrt{-\tilde{g}}F(R),
\end{equation}
where $F(R)$ is a double differentiable function of the Ricci scalar $R$. Introducing a new scalar field $\sigma$ without the kinetic term, we rewrite $S_{R}$ as follows~\cite{Maeda:1988ab,Kaneda:2015jma}:
\begin{equation}
    \tilde{S}_{J} =\int d^4 \tilde{x} \sqrt{-\tilde{g}} \left[\frac{\partial F(\sigma)}{\partial\sigma}(R-\sigma)+ F(\sigma)\right].
\end{equation}

By the conformal transformation of the metric
$g_{\mu\nu}=\frac{2f(\sigma)}{M^2_{Pl}}\tilde{g}_{\mu\nu}$,
where $f\equiv\frac{d F(\sigma)}{d\sigma}$,
one gets the following action in the Einstein frame~\cite{SBB1989}:
\begin{equation}
\label{SE}
S_{E}=\int d^4x\sqrt{-g}\left[\frac{M^2_{Pl}}{2}R_E-\frac{h(\sigma)}{2}{g^{\mu\nu}}\partial_\mu{\sigma}\partial_\nu{\sigma}-V_E\right],
\end{equation}
where
\begin{equation*}
%\label{potW}
h(\sigma)=\frac{3M^2_{Pl}}{2f^2}\left(\frac{d f}{d\sigma}\right)^2,\qquad V_E= M^4_{Pl}\frac{f\sigma-F}{4f^2} \,.
\end{equation*}

Introducing the scalar field
\begin{equation}
\psi=\sqrt{\frac{3}{2}}M_{Pl}\ln\left(\frac{2}{M^2_{Pl}}f(\sigma)\right),
\end{equation}
 we obtain the action $S_{E}$  as follows:
\begin{equation}\label{ActionSe}
    S_E=\int  d^4x \sqrt{-g}\left[\frac{ M^2_{Pl}}{2}R_E-\frac12\partial_\mu\psi\partial_\mu\psi-V_E(\psi)\right].
\end{equation}
%The standard choice is $f_0=M_{Pl}^2/2$.

So, we get the Einstein frame model with a standard scalar field. To obtain inverse transformation we present the potential and its derivative in the following form:
\begin{equation*}
V_E(\psi) =  \frac{M^2_{Pl}}{2} R\, \mathrm{e}^{ - \frac{\sqrt{6}\psi}{3M_{Pl}}} -F\, \mathrm{e}^{- \frac{2\sqrt{6}\psi}{3M_{Pl}}},\qquad
\frac{dV_E(\psi)}{d \psi} = {}-\frac{M_{Pl}}{\sqrt{6}}R\, \mathrm{e}^{ -\frac{\sqrt{6}\psi}{3M_{Pl}}} + \frac{4}{\sqrt{6}M_{Pl}}F\, \mathrm{e}^{- \frac{2\sqrt{6}\psi}{3M_{Pl}}}.
\end{equation*}

So, we get the function $F(R)$ in a parametric form~\cite{Ketov:2014jta,Motohashi:2017vdc,Ketov:2019toi}:
\begin{eqnarray}
R &=& \left[\frac{\sqrt{6}}{M_{Pl}}\frac{dV_E}{d\psi}+\frac{4V_E}{M^2_{Pl}}\right]\mathrm{e}^{\frac{\sqrt{6}\psi}{{3}M_{Pl}}},
\label{RV}\\
F &=& \frac{M^2_{Pl}}{2}\left[\frac{\sqrt{6}}{M_{Pl}}\frac{dV_E}{d\psi}+\frac{2V_E}{M^2_{Pl}}\right]\mathrm{e}^{2\frac{\sqrt{6}\psi}{{3}M_{Pl}}}.
\label{FV}
\end{eqnarray}

So, if model with one minimally coupled scalar field has exact solutions, then the corre\-sponding  $F(R)$ gravity model has them as well.
The goal of this paper is to find  such potentials $V_E$ that the Einstein frame model has exact solutions and the function $F(R)$ can be found in the analytic form.

\section{Construction of $F(R)$ gravity models}

If Eq.~(\ref{RV}) has the solution
\begin{equation}
\label{Rform}
R = C_1 + C_k \mathrm{e}^{ \frac{k\psi}{M_{Pl}}},
\end{equation}
where $C_1$ and $C_k\neq 0$ are arbitrary constants, then the function $F(R)$ can be obtained in the analytic form.
From Eqs.~(\ref{RV}) and (\ref{Rform}), we get the following linear first order differential equation for the potential $V_E$:
\begin{equation}
\label{EquV}
\left[\frac{\sqrt{6}}{M_{Pl}}\frac{dV_E}{d\psi}+\frac{4V_E}{M^2_{Pl}}\right]\mathrm{e}^{\frac{\sqrt{6}}{3}\frac{\psi}{M_{Pl}}} =  C_1 + C_k \mathrm{e}^{k \frac{\psi}{M_{Pl}}}.
\end{equation}
Equation (\ref{EquV}) has the general solution:
\begin{equation}
\label{Vform}
V_E(\psi) =  \frac{M_{Pl}^2}{2}\left(C_2 \mathrm{e}^{-2\frac{\sqrt{6}\psi}{{3}M_{Pl}}} + C_1 \mathrm{e}^{-\frac{\sqrt{6}\psi}{{3}M_{Pl}}} + C_{\omega} \mathrm{e}^{\omega\frac{\sqrt{6}\psi}{{3}M_{Pl}}}\right),
\end{equation}
where $C_2$ is an integration constant, \  $\omega=\sqrt{6}k/2-1$, \ and \
$ C_\omega=\frac{\sqrt{6}C_k}{\sqrt{6}+3k}=\frac{C_k}{\omega+2}$.

Substituting the potential (\ref{Vform}) into (\ref{RV}) and (\ref{FV}), we obtain the following expressions:
\begin{equation}
\label{RFpsi}
R =
C_\omega(\omega + 2) \mathrm{e}^{(\omega + 1)\frac{\sqrt{6}\psi}{{3}M_{Pl}}} + C_1,
\qquad
F=\frac{M^2_{Pl}}{2}\left(
C_\omega(\omega +  1) \mathrm{e}^{(\omega + 2)\frac{\sqrt{6}\psi}{{3}M_{Pl}}} - C_2\right).
\end{equation}

Finally, we get
\begin{equation}
\label{FR}
F(R) = \frac{M^2_{Pl}}{2}\left(C_\omega(\omega + 1)\left( \frac{R - C_1}{C_\omega (\omega + 2)}\right)^\alpha - C_2\right), \quad \mbox{where}\quad \alpha =\frac{\omega + 2}{\omega+1}.
\end{equation}
It is easy to see that  $\alpha\neq 1$ for any $\omega$, also $\alpha=2$ corresponds to $\omega=0$.

\section{The search of exact solutions}

Let us consider the potential (\ref{Vform})  in the case of $C_1=C_2=0$. If $\omega\neq 0$, then we get an exponential potential and an integrable cosmological model~\cite{SalopekBond,gen-exp,Fre,KPTVV2013}. The general solutions of this model can be found explicitly in a parametric time~\cite{Fre}.
The general solutions of the corresponding $R^\alpha$ models with an arbitrary $\alpha$, but $\alpha\neq 2$ and $\alpha\neq 1$, can be obtained from the general solution of the model with an exponential potential by the conformal transformation of the metric.
In the case of $\omega= 0$, we get the model with the cosmological constant that is integrable (the general solution is presented in~\cite{Aref'eva:2007yr,Kamenshchik:2016}) and corresponds to a pure $R^2$ gravity model.

To prove the integrability of the cosmological model with an exponential potential the superpotential method has been used in the paper~\cite{SalopekBond}. This method is actively used to get cosmological models with exact particular solutions both with one scalar field~\cite{SalopekBond,Muslimov,Townsend,AKV,Bazeia,Chervon:2017kgn,
KTVV2013} and with a few scalar fields~\cite{SalopekBond,AKV2,Andrianov:2007ua,CFPSV2019}, as well as to construct inflationary models~\cite{Lidsey:1995np,Chervon:2008zz,Vennin:2014xta,Binetruy:2014zya}. We use this method to get models with exact solutions in the cases of $C_1^2+C^2_2\neq 0$.

For the spatially flat Friedmann--Lema\^{i}tre--Robertson--Walker metric with
\begin{equation} \label{FLRW}
	{ds}^{2}={}-{dt}^2+a_E^2(t)\left(dx_1^2+dx_2^2+dx_3^2\right),
\end{equation}
the Einstein equations can be written in the following form:
\begin{equation}
	\label{trequ} \dot\psi={}-2M_{Pl}^2W'_{,\psi} \,,
\end{equation}
\begin{equation}
	\label{a2} V_E=3M_{Pl}^2 W^2 -2M^4_{Pl}{W'_{,\psi}}^2\,,
\end{equation}
where the Hubble parameter  $H_E\equiv\dot{a}/a=W(\psi)$ and $W'_{,\psi}=\frac{dW}{d\psi}$.

Choosing the superpotential
\begin{equation}
W(\psi) = W_a \mathrm{e}^{a\frac{\sqrt{6}\psi}{{2}M_{Pl}}} + W_b \mathrm{e}^{b\frac{\sqrt{6}\psi}{{2}M_{Pl}}},
\end{equation}
where $a$, $b$, $W_a$ and $W_b$ are constants, we get the following potential:
\begin{equation}
\label{ab_eqn}
V_E= 6\left(W_a^2(1-a^2)\mathrm{e}^{a\sqrt{6} \frac{\psi}{M_{Pl}}} + 2W_a W_b (1 - ab)\mathrm{e}^{\frac{(a+b)}{2}\sqrt{6} \frac{\psi}{M_{Pl}}} +  W_b^2(1-b^2)\mathrm{e}^{b\sqrt{6} \frac{\psi}{M_{Pl}}}\right).
\end{equation}

For some values of parameters $a$ and $b$, we get potentials in the form~(\ref{Vform}) and the corresponding $F(R)$ gravity models with exact solutions (see Table~\ref{LisT}). We can assume that $a<b$ without loss of generality.
In the cases $a = -1/3$, $b = 1$ and $a = -1$ , $b = 1/3$ the resulting $F(R)$ models are coincide. In the general case the particular solution $\psi(t)$ can be obtained in quadratures by integrating Eq.~(\ref{trequ}).

\begin{table}[h]
\begin{center}
\caption{List of the $F(R)$ models obtained.}
\begin{tabular}{|c|c|c|c|c|c|c|}
\hline
$a$, $b$ & $\omega$ & $\alpha$ & $C_2$ & $C_1$ & $C_\omega$ & $2F(R)/M^2_{Pl}\vphantom{\Big(}$ \\[1mm]
\hline
$a = -\frac{2}{3}, b = 0$ & $0$ & $2$ & $\frac{10}{3}W_a^2$ & $12W_aW_b$ & $6W_b^2$ & $\frac{1}{24W_b^2}R^2 - \frac{W_a}{W_b}R + \frac{8W_a^2}{3}$ \\
\hline
$a = -\frac{2}{3}, b = 1$ & $\frac{1}{2}$ & $\frac{5}{3}$ & $\frac{10}{3}W_a^2$ & $0$ & $20W_aW_b$ & $30W_aW_b\left( \frac{R}{50W_aW_b} \right)^{5/3} - \frac{10W_a^2}{3}$ \\
\hline
$a = -1, b = \frac{1}{3}$ & $1$ & $\frac{3}{2}$ & $0$ & $16W_a W_b$ & $\frac{16}{3}W_b^2$ & $\frac{32}{3}W_b^2\left(\frac{1}{16W_b^2}R - \frac{W_a}{W_b}\right)^{3/2}$ \\
\hline
$a = -\frac{1}{3}, b = 1$ & $1$ & $\frac{3}{2}$ & $0$ & $\frac{16}{3}W_a^2$ & $16W_aW_b$ & $32W_aW_b\left( \frac{R}{48W_aW_b} - \frac{W_a}{9W_b} \right)^{3/2}$ \\
\hline
$a = -3, b = -\frac{1}{3}$ & $-9$ & $\frac{7}{8}$ & $0$ & $\frac{16}{3}W_b^2$ & $-48W_a^2$ & $384W_a^2\left( \frac{1}{336W_a^2}R - \frac{W_b^2}{63W_a^2}\right)^{7/8} $ \\
\hline
$a = -\frac{7}{3}, b = 1$ & $-7$ & $\frac{5}{6}$ & $40 W_a W_b$ & $0$ & $-\frac{80}{3}W_a^2$ & $160W_a^2\left[\frac{3R}{400W_a^2}\right]^{5/6} - 40W_aW_b\! $ \\
\hline
$a = -\frac{5}{3}, b = 1$ & $-5$ & $\frac{3}{4}$ & $0$ & $32 W_a W_b$ & $-\frac{32}{3}W_a^2$ & $ \frac{128}{3}W_a^2 \left( \frac{R}{32W_a^2} - \frac{W_b}{W_a} \right)^{3/4} $ \\
\hline
$a = -\frac{3}{2}, b = -\frac{2}{3}$ & $-\frac{9}{2}$ & $\frac{5}{7}$ & $\frac{10}{3}W_b^2$ & $0$ & $-\frac{15}{2}W_a^2$ & $\frac{105}{4}W_a^2\left(\frac{4}{75 W_a^2}R\right)^{5/7} - \frac{10W_b^2}{3}$ \\
\hline
$a = -1, b = -\frac{2}{3}$ & $-\frac{5}{2}$ & $\frac{1}{3}$ & $\frac{10}{3}W_b^2$ & $0$ & $4W_aW_b$ & $6 W_a W_b\left(\frac{1}{2W_a W_b}R\right)^{1/3} - \frac{10W_b^2}{3}$ \\
\hline
$a = -\frac{2}{3}, b = -\frac{1}{3}$ & $-\frac{3}{2}$ & $-1\!$ & $\frac{10}{3}W_a^2$ & $\frac{16}{3}W_b^2$ & $\frac{28}{3}W_a W_b$ & $\frac{196W_a^2W_b^2}{3(3R - 16W_b^2)} - \frac{10W_a^2}{3}$ \\
\hline
\end{tabular}
\label{LisT}
\end{center}
\end{table}

\section{The case of the $R^2$ gravity}

Let us consider in detail the case of the $R^2$ gravity model with
\begin{equation}\label{FR2}
    F(R)=\frac{M^2_{Pl}}{2}\left(\frac{1}{24W_b^2}R^2 - \frac{W_a}{W_b}R + \frac{8W_a^2}{3}\right),
\end{equation}
that corresponds to $W(\psi)=W_a \exp\left({}-\frac{\sqrt{6}\psi}{{3}M_{Pl}}\right)+W_b $.

Equation~(\ref{trequ}) leads to
\begin{equation}
\frac{d\psi}{dt}=\frac{2\sqrt {6}}{3}M_{Pl} W_a\,{\mathrm{e}^{-{\frac{\sqrt {6}\psi}{3M_{Pl}}}}},\quad\Rightarrow\quad \psi=\frac{\sqrt {6}}{2}M_{Pl}\ln \left(\frac{4W_a}{3}\left(t-t_0\right) \right),
\end{equation}
where $t_0$ is an integration constant.

Substituting $\psi(t)$ into  $W(\psi)$, we get the Hubble parameter for the model with the scalar field:
\begin{equation}
H_E=\frac{3}{4(t-t_0)}+W_b.
\end{equation}

In the initial $F(R)$ model, we get the Friedmann--Lema\^{i}tre--Robertson--Walker metric with parametric time $t$ and
\begin{equation}
d\tilde{s}^2={}-\frac{M^2_{Pl}}{2f(R)}{d{t}}^2+\tilde{a}^2\left(d{x}_1^2+d{x}_2^2+d{x}_3^2\right), \quad\mbox{where}\quad \tilde{a}^2=\frac{M^2_{Pl}}{2f(R)}{a}_E^2.
\end{equation}

Using Eq.~(\ref{RFpsi}), we get
\begin{equation*}
f(R)=\frac{M^2_{Pl}}{2}\exp\left(\frac{\sqrt{6}}{3}\frac{\psi}{M_{Pl}}\right)=\frac{2}{3}M^2_{Pl}W_a(t-t_0).
\end{equation*}

The cosmic time in this frame is
\begin{equation}
\label{time_transformation}
\tilde{t}=\int\sqrt{\frac{M^2_{Pl}}{2f(\sigma)}}{d{t}}=\sqrt{\frac{3(t-t_0)}{W_a}}+\tilde{t}_0.
\end{equation}

The corresponding Hubble parameter can be presented in the form:
\begin{equation}\label{tildeH}
\tilde{H}=\tilde{a}^{-1}\frac{d\tilde{a}}{d\tilde{t}}=\sqrt{\frac{2f(R)}{M^2_{Pl}}}\left[H_E-\frac12\frac{d\ln(f)}{dt}\right]=
\frac{1}{2(\tilde{t}-\tilde{t}_0)}+\frac{2}{3}W_bW_a(\tilde{t}-\tilde{t}_0).
\end{equation}
The first term of this expression corresponds to the radiation dominated universe, whereas the second term is the Ruzmaikina--Rusmaikin solution~\cite{Ruzmaikina}.

\section{Conclusions}

In this paper, we have found a few $F(R)$ gravity models with exact solutions and shown that the superpotential method is a useful tool for this propose.
The existence of a fundamental scalar field (the Higgs boson) gives good motivation to consider modified gravity models with an additional scalar field.
The $F(R,\chi)$ gravity models with the scalar field $\chi$~\cite{Gottlober:1993hp} are very popular~\cite{Kaneda:2015jma,delaCruz-Dombriz:2016bjj,Wang:2017fuy,He:2018gyf,Gorbunov:2018llf,Karam:2018mft} as models of inflation, in particular, the mixed Higgs$-R^2$ model~\cite{Wang:2017fuy,He:2018gyf,Gorbunov:2018llf}. We plan to generalize the investigation on the $F(R,\chi)$ models and to use the superpotential method developed for the search of exact solutions of the chiral cosmological models~\cite{CFPSV2019}, or some other methods~\cite{Paliathanasis} to construct physically interesting $F(R,\chi)$ models with exact solutions.

E.O.P. and S.Yu.V. are supported in part by RFBR, project~18-52-45016.


\begin{thebibliography}{99}


\bibitem{Sotiriou:2008rp}
  \textit{T.P.~Sotiriou and V.~Faraoni,}
  $f(R)$ Theories of Gravity,
  Rev.\ Mod.\ Phys.\  {\bf 82} (2010) 451, arXiv:0805.1726
 % doi:10.1103/RevModPhys.82.451
   %%CITATION = doi:10.1103/RevModPhys.82.451;%%

 \bibitem{Tsujikawa:2010}
 \textit{A.~De Felice and S.~Tsujikawa},
  $f(R)$ theories,
  Living Rev.\ Rel.\  {\bf 13} (2010) 3, arXiv:1002.4928
 % doi:10.12942/lrr-2010-3

\bibitem{Faulkner:2006ub}
\textit{T.~Faulkner, M.~Tegmark, E.F.~Bunn and Y.~Mao},
  Constraining $f(R)$ Gravity as a Scalar Tensor Theory,
  Phys.\ Rev.\ D {\bf 76} (2007) 063505, arXiv:astro-ph/0612569
  %doi:10.1103/PhysRevD.76.063505
    %%CITATION = doi:10.1103/PhysRevD.76.063505;%%

\bibitem{Starobinsky}
 \textit{A.A.~Starobinsky},
   A New Type of Isotropic Cosmological Models Without Singularity,
    Phys.\ Lett. B {\bf 91} (1980)  99;\\
\textit{A.A.~Starobinsky},
		Dynamics of phase transition in the new inflationary universe scenario and generation of perturbations,
		Phys. Lett. B {\bf 117} (1982) 175.



\bibitem{Mijic:1986iv}
  \textit{M.B.~Mijic, M.S.~Morris and W.M.~Suen},
  The R**2 Cosmology: Inflation Without a Phase Transition,
  Phys.\ Rev.\ D {\bf 34} (1986) 2934;\\
 % doi:10.1103/PhysRevD.34.2934
  %%CITATION = doi:10.1103/PhysRevD.34.2934;%%
 \textit{ K.~Maeda},
  Inflation as a Transient Attractor in R**2 Cosmology,
  Phys.\ Rev.\ D {\bf 37} (1988) 858
 % doi:10.1103/PhysRevD.37.858
  %%CITATION = doi:10.1103/PhysRevD.37.858;%%

\bibitem{Planck2018}
   \textit{Y.~Akrami {\it et al.}} [Planck Collaboration],
  Planck 2018 results. X. Constraints on inflation,
  arXiv:1807.06211
  %%CITATION = ARXIV:1807.06211;%%
  %%CITATION = ARXIV:1807.06209;%%

\bibitem{Capozziello:2003gx}
  \textit{S.~Capozziello, V.~F.~Cardone, S.~Carloni and A.~Troisi},
  Curvature quintessence matched with observational data,
  Int.\ J.\ Mod.\ Phys.\ D {\bf 12} (2003) 1969, arXiv:astro-ph/0307018
 % doi:10.1142/S0218271803004407
   %%CITATION = doi:10.1142/S0218271803004407;%%


\bibitem{Dolgov:2003px}
  \textit{A.D.~Dolgov and M.~Kawasaki},
  Can modified gravity explain accelerated cosmic expansion?,
  Phys.\ Lett.\ B {\bf 573} (2003) 1, arXiv:astro-ph/0307285
  %doi:10.1016/j.physletb.2003.08.039
  %%CITATION = doi:10.1016/j.physletb.2003.08.039;%%


\bibitem{Hu:2007nk}
  \textit{W.~Hu and I.~Sawicki},
  Models of $f(R)$ Cosmic Acceleration that Evade Solar-System Tests,
  Phys.\ Rev.\ D {\bf 76} (2007) 064004, arXiv:0705.1158;\\
 % doi:10.1103/PhysRevD.76.064004
  %%CITATION = doi:10.1103/PhysRevD.76.064004;%%
\textit{K.~Bamba, C.~Q.~Geng, S.~Nojiri and S.~D.~Odintsov},
Crossing of the phantom divide in modified gravity,
  Phys.\ Rev.\ D {\bf 79} (2009) 083014, arXiv:0810.4296
  %%CITATION = doi:10.1103/PhysRevD.79.083014;%%

 \bibitem{Starobinsky:2007hu}
  \textit{A.A.~Starobinsky},
  Disappearing cosmological constant in $f(R)$ gravity,
  JETP Lett.\  {\bf 86} (2007) 157, arXiv:0706.2041
  %doi:10.1134/S0021364007150027
    %%CITATION = doi:10.1134/S0021364007150027;%%

 \bibitem{Tsujikawa:2007xu}
  \textit{S.~Tsujikawa},
  Observational signatures of $f(R)$ dark energy models that satisfy cosmological and local gravity constraints,
  Phys.\ Rev.\ D {\bf 77} (2008) 023507,
 % doi:10.1103/PhysRevD.77.023507
  arXiv:0709.1391
  %%CITATION = doi:10.1103/PhysRevD.77.023507;%%

\bibitem{Bamba:2008hq}
  \textit{K.~Bamba, C.Q.~Geng, S.~Nojiri and S.D.~Odintsov},
  Crossing of the phantom divide in modified gravity,
  Phys.\ Rev.\ D {\bf 79} (2009) 083014,
  %doi:10.1103/PhysRevD.79.083014
  arXiv:0810.4296
  %%CITATION = doi:10.1103/PhysRevD.79.083014;%%

\bibitem{Ali:2010zx}
  \textit{A.~Ali, R.~Gannouji, M.~Sami and A.~A.~Sen},
  Background cosmological dynamics in $f(R)$ gravity and observational constraints,
  Phys.\ Rev.\ D {\bf 81} (2010) 104029,
 % doi:10.1103/PhysRevD.81.104029
  arXiv:1001.5384
  %%CITATION = doi:10.1103/PhysRevD.81.104029;%%

 \bibitem{Capozziello:2018wul}
  \textit{S.~Capozziello, S.~Nojiri and S.~D.~Odintsov},
 The role of energy conditions in $f(R)$ cosmology,
  Phys.\ Lett.\ B {\bf 781} (2018) 99, arXiv:1803.08815
 % doi:10.1016/j.physletb.2018.03.064
  %%CITATION = doi:10.1016/j.physletb.2018.03.064;%%

\bibitem{Arbuzova:2019ptx}
  \textit{E.~Arbuzova},
  Instabilities in modified theories of gravity,
  arXiv:1911.02892
  %%CITATION = ARXIV:1911.02892;%%




  \bibitem{Nojiri:2003ft}
  \textit{S.~Nojiri and S.~D.~Odintsov},
  Modified gravity with negative and positive powers of the curvature: Unification of the inflation and of the cosmic acceleration,
  Phys.\ Rev.\ D {\bf 68}  (2003) 123512, arXiv:hep-th/0307288;\\
  %doi:10.1103/PhysRevD.68.123512
    %%CITATION = doi:10.1103/PhysRevD.68.123512;%%
  \textit{G.~Cognola, E.~Elizalde, S.~Nojiri, S.D.~Odintsov, L.~Sebastiani and S.~Zerbini},
  A Class of viable modified $f(R)$ gravities describing inflation and the onset of accelerated expansion,
  Phys.\ Rev.\ D {\bf 77} (2008) 046009, arXiv:0712.4017
 % doi:10.1103/PhysRevD.77.046009
    %%CITATION = doi:10.1103/PhysRevD.77.046009;%%



\bibitem{Motohashi:2010tb}
 \textit{H.~Motohashi, A.A.~Starobinsky and J.~Yokoyama},
  Phantom boundary crossing and anomalous growth index of fluctuations in viable $f(R)$ models of cosmic acceleration,
  Prog.\ Theor.\ Phys.\  {\bf 123} (2010) 887, arXiv:1002.1141
%  doi:10.1143/PTP.123.887
  %%CITATION = doi:10.1143/PTP.123.887;%%

\bibitem{OdintsovRev}
\textit{S.~Nojiri and S.~D.~Odintsov},
Unified cosmic history in modified gravity: from $F(R)$ theory to Lorentz non-invariant models,
  Phys.\ Rept.\  {\bf 505} (2011) 59,
%  doi:10.1016/j.physrep.2011.04.001
arXiv:1011.0544;\\
  %%CITATION = doi:10.1016/j.physrep.2011.04.001;%%
\textit{S.~Nojiri, S.D.~Odintsov and V.K.~Oikonomou},
Modified Gravity Theories on a Nutshell: Inflation, Bounce and Late-time Evolution,
  Phys.\ Rept.\  {\bf 692} (2017) 1, arXiv:1705.11098
  %%CITATION = doi:10.1016/j.physrep.2017.06.001;%%


\bibitem{Paliathanasis:2017apr}
  \textit{A.~Paliathanasis},
  Analytic Solution of the Starobinsky Model for Inflation,
  Eur.\ Phys.\ J.\ C {\bf 77} (2017) 438, arXiv:1706.06400;\\
  %doi:10.1140/epjc/s10052-017-5009-0
  %%CITATION = doi:10.1140/epjc/s10052-017-5009-0;%%
\textit{G.~Papagiannopoulos, S.~Basilakos, J.~D.~Barrow and A.~Paliathanasis},
  New integrable models and analytical solutions in $f(R)$ cosmology with an ideal gas,
  Phys.\ Rev.\ D {\bf 97} (2018) 024026, arXiv:1801.01274
  %doi:10.1103/PhysRevD.97.024026
  %%CITATION = doi:10.1103/PhysRevD.97.024026;%%

\bibitem{Muller:2017nxg}
  \textit{D.~Muller, A.~Ricciardone, A.~A.~Starobinsky and A.~Toporensky},
  Anisotropic cosmological solutions in $R + R^2$ gravity,
  Eur.\ Phys.\ J.\ C {\bf 78} (2018) 311, arXiv:1710.08753
  %doi:10.1140/epjc/s10052-018-5778-0
  %%CITATION = doi:10.1140/epjc/s10052-018-5778-0;%%



\bibitem{Maeda:1988ab}
  \textit{K.i.~Maeda},
  {Towards the Einstein-Hilbert Action via Conformal Transformation},
  Phys.\ Rev.\ D {\bf 39} (1989) 3159
  %doi:10.1103/PhysRevD.39.3159
  %%CITATION = doi:10.1103/PhysRevD.39.3159;%%
	
  \bibitem{SalopekBond}
\textit{D.S.~Salopek and J.R.~Bond},
Nonlinear evolution of long-wavelength metric fluctuations in inflationary models,
Phys. Rev. D \textbf{42} (1990) 3936

\bibitem{Muslimov} \textit{A.G.~Muslimov}, On the Scalar Field Dynamics
		in a Spatially Flat Friedman Universe, Class. Quant. Grav. \textbf{7}
		(1990) 231

\bibitem{Kaneda:2015jma}
  \textit{S.~Kaneda and S.V.~Ketov},
 Starobinsky-like two-field inflation,
  Eur.\ Phys.\ J.\ C {\bf 76} (2016) no.1,  26,
 % doi:10.1140/epjc/s10052-016-3888-0
  arXiv:1510.03524
  %%CITATION = doi:10.1140/epjc/s10052-016-3888-0;%%


\bibitem{SBB1989}
\textit{D.S.~Salopek, J.R.~Bond and J.M.~Bardeen},
Designing Density Fluctuation Spectra in Inflation,
 Phys. Rev. D \textbf{40} (1989) 1753

\bibitem{Ketov:2014jta}
\textit{S.V.~Ketov and N.~Watanabe},
The $f(R)$ gravity function of Linde quintessence,
{Phys. Lett. B} {\bf 741} (2015) 242, arXiv:1410.3557


\bibitem{Motohashi:2017vdc}
  \textit{H.~Motohashi and A.A.~Starobinsky},
  $f(R)$ constant-roll inflation,
  Eur.\ Phys.\ J.\ C {\bf 77}, no. 8 (2017) 538,
  %doi:10.1140/epjc/s10052-017-5109-x
  arXiv:1704.08188
  %%CITATION = doi:10.1140/epjc/s10052-017-5109-x;%%

\bibitem{Ketov:2019toi}
  \textit{S.V.~Ketov},
  On the equivalence between Starobinsky and Higgs inflationary models in gravity and supergravity,
  arXiv:1911.01008.
  %%CITATION = ARXIV:1911.01008;%%

\bibitem{gen-exp}
\textit{V.~Muller, H.J.~Schmidt and A.A.~Starobinsky},
  Power law inflation as an attractor solution for inhomogeneous cosmological models,
  Class.\ Quant.\ Grav.\  {\bf 7} (1990) 1163;\\
 \textit{E.~Elizalde, S.~Nojiri and S.D.~Odintsov},
  Late-time cosmology in (phantom) scalar-tensor theory: Dark energy and the cosmic speed-up,
  Phys.\ Rev.\ D {\bf 70} (2004) 043539;
  arXiv:hep-th/0405034; \\
  %%CITATION = doi:10.1103/PhysRevD.70.043539;%%
 \textit{A.A.~Andrianov, F.~Cannata and A.Y.~Kamenshchik},
  General solution of scalar field cosmology with a (piecewise) exponential potential,
  J. Cosmol. Astropart. Phys. {\bf 1110} (2011) 004, arXiv:1105.4515
  %%CITATION = doi:10.1088/1475-7516/2011/10/004;%%

\bibitem{Fre}
 \textit{P.~Fr\'e, A.~Sagnotti, A.S.~Sorin},
Integrable Scalar Cosmologies I. Foundations and links with String Theory,
Nucl. Phys. B {\bf 877} (2013)  1028, arXiv:1307.1910

\bibitem{KPTVV2013}
\textit{A.Yu.~Kamenshchik, E.O.~Pozdeeva, A.~Tronconi, G.~Venturi, S.Yu.~Vernov},
Integrable cosmological models with non-minimally coupled scalar fields,
  Class.\ Quant.\ Grav.\  {\bf 31} (2014)  105003, arXiv:1307.1910

\bibitem{Aref'eva:2007yr}
  \textit{I.Ya.~Aref'eva, L.V.~Joukovskaya, S.Yu.~Vernov},
  Dynamics in nonlocal linear models in the Friedmann-Robertson-Walker metric,
  J.\ Phys.\ A {\bf 41} (2008) 304003, arXiv:0711.1364

\bibitem{Kamenshchik:2016}
  \textit{A.Yu.~Kamenshchik, E.O.~Pozdeeva, S.Yu.~Vernov, A.~Tronconi and G.~Venturi},
Transformations between Jordan and Einstein frames: Bounces, antigravity, and crossing singularities,
  Phys.\ Rev.\ D {\bf 94} (2016)  063510, arXiv:1602.07192;\\
 % doi:10.1103/PhysRevD.94.063510
  %%CITATION = doi:10.1103/PhysRevD.94.063510;%%
  \textit{A.Yu.~Kamenshchik, E.O.~Pozdeeva, A.~Tronconi, G.~Venturi and S.Yu.~Vernov},
  General solutions of integrable cosmological models with non-minimal coupling,
  Phys.\ Part.\ Nucl.\ Lett.\  {\bf 14}, no. 2  (2017) 382, arXiv:1604.01959;\\
  %doi:10.1134/S1547477117020169
   %%CITATION = doi:10.1134/S1547477117020169;%%
  \textit{A.Yu.~Kamenshchik, E.O.~Pozdeeva, A.~Tronconi, G.~Venturi and S.Yu.~Vernov},
  Integrable cosmological models in the Einstein and in the Jordan frames and Bianchi-I cosmology,
  Phys.\ Part.\ Nucl.\  {\bf 49}, no. 1 (2018) 1, arXiv:1606.04260
 % doi:10.1134/S1063779618010173
  %%CITATION = doi:10.1134/S1063779618010173;%%
		
		\bibitem{Townsend}
		\textit{K. Skenderis and P.K. Townsend}, Hamilton-Jacobi method for
		Domain Walls and Cosmologies, Phys. Rev. D \textbf{74} (2006) 125008, arXiv:hep-th/0609056;\\
	  \textit{P.K. Townsend}, Hamilton-Jacobi Mechanics from
		Pseudo-Supersymmetry, Class. Quant. Grav. \textbf{25} (2008) 045017, arXiv:0710.5178

		\bibitem{AKV} \textit{I.Ya.~Aref'eva,
		A.S.~Koshelev, and S.Yu.~Vernov}, Exactly Solvable SFT Inspired
		Phantom Model, Theor. Math. Phys. \textbf{148} (2006) 895, arXiv:astro-ph/0412619
%%CITATION = ASTRO-PH 0412619;%%

\bibitem{Bazeia}
\textit{D. Bazeia, C.B. Gomes, L. Losano, and R. Menezes}, First-order
		formalism and dark energy, Phys. Lett. B \textbf{633} (2006) 415, arXiv:astro-ph/0512197;\\
\textit{D. Bazeia, L. Losano, R. Rosenfeld},
First-order formalism for dust,
Eur. Phys. J. C \textbf{55} (2008) 113, arXiv:astro-ph/0611770


\bibitem{Chervon:2017kgn}
\textit{S.V.~Chervon, I.V. Fomin and A. Beesham},
 The method of generating functions in exact scalar field cosmology,
 Eur.\ Phys.\ J.\ C {\bf 78} (2018)   301, arXiv:1704.08712;\\
 % doi:10.1140/epjc/s10052-018-5795-z
  %%CITATION = doi:10.1140/epjc/s10052-018-5795-z;%%
\textit{{T.~Harko, F.S.N.~Lobo, and M.K.~Mak}},
Arbitrary scalar field and quintessence cosmological models,
  Eur.\ Phys.\ J.\ C {\bf 74} (2014) 2784, arXiv:1310.7167
 % doi:10.1140/epjc/s10052-014-2784-8
  %%CITATION = doi:10.1140/epjc/s10052-014-2784-8;%%


\bibitem{KTVV2013}
\textit{A.Yu.~Kamenshchik, A.~Tronconi, G.~Venturi, and S.Yu.~Vernov},
Reconstruction of Scalar Potentials in Modified Gravity Models,
Phys. Rev. D \textbf{87} (2013) 063503, arXiv:1211.6272

\bibitem{AKV2} \textit{I.Ya.~Aref'eva, A.S.~Koshelev, and S.Yu~Vernov}, Crossing the $w=-1$ barrier
		in the D3-brane dark energy model,  Phys. Rev. D \textbf{72} (2005)
		064017, arXiv:astro-ph/0507067;\\
%%CITATION = ASTRO-PH 0507067;%%
 \textit{S.Yu~Vernov},
		Construction of Exact Solutions in Two-Field Models,
		Theor. Math. Phys. \textbf{155} (2008) 544, arXiv:astro-ph/0612487;\\
%%CITATION = ASTRO-PH 0612487;%%
\textit{I.Ya.~Aref'eva, N.V.~Bulatov and S.Yu.~Vernov},
		Stable Exact Solutions in Cosmological Models with Two Scalar Fields,
		Theor.\ Math.\ Phys.\  {\bf 163} (2010) 788, arXiv:0911.5105
		%doi:10.1007/s11232-010-0063-x
		  %%CITATION = doi:10.1007/s11232-010-0063-x;%%


\bibitem{Andrianov:2007ua}
 \textit{{A.A.~Andrianov, F.~Cannata, A.Yu.~Kamenshchik, and D.~Regoli}},
Reconstruction of scalar potentials in two-field cosmological models,
J. Cosmol. Astropart. Phys. {\bf 0802} (2008) 015, arXiv:0711.4300;\\
 \textit{M.R.~Setare, J.~Sadeghi},
First-order formalism for the quintom model of dark energy,
   	Int. J. Theor. Phys. \textbf{47} (2008) 3219, arXiv:0805.1117

\bibitem{CFPSV2019}
\textit{S.V.~Chervon, I.V.~Fomin, E.O.~Pozdeeva, M.~Sami and S.Yu.~Vernov},
  Superpotential method for chiral cosmological models connected with modified gravity,
  Phys.\ Rev.\ D {\bf 100} (2019)  063522, arXiv:1904.11264
  %doi:10.1103/PhysRevD.100.063522
  %%CITATION = doi:10.1103/PhysRevD.100.063522;%%


\bibitem{Lidsey:1995np}
  \textit{J.E.~Lidsey, A.R.~Liddle, E.W.~Kolb, E.J.~Copeland, T.~Barreiro and M.~Abney},
  Reconstructing the inflation potential: An overview,
  Rev. Mod. Phys. \textbf{69} (1997) 373, arXiv:astro-ph/9508078
   %%CITATION = doi:10.1103/RevModPhys.69.373;%%


 \bibitem{Chervon:2008zz}
 \textit{S.V.~Chervon and I.V.~Fomin},
 On calculation of the cosmological parameters in exact models of inflation,
  Grav.\ Cosmol.\  {\bf 14} (2008) 163, arXiv:1704.05378;\\
  %doi:10.1134/S0202289308020060
\textit{A.V.~Yurov, V.A.~Yurov, S.V.~Chervon and M.~Sami}, Potential of total energy as superpotential in integrable cosmological models, Theor. Math. Phys. \textbf{166} (2011) 259.

\bibitem{Vennin:2014xta}
  \textit{V.~Vennin},
  Horizon-Flow off-track for Inflation,
  Phys.\ Rev.\ D {\bf 89} (2014)  083526, arXiv:1401.2926
  %doi:10.1103/PhysRevD.89.083526
 % [arXiv:1401.2926 [astro-ph.CO]].
  %%CITATION = doi:10.1103/PhysRevD.89.083526;%%

\bibitem{Binetruy:2014zya}
		\textit{P.~Binetruy, E.~Kiritsis, J.~Mabillard, M.~Pieroni and C.~Rosset,}
		Universality classes for models of inflation,
		J. Cosmol. Astropart. Phys.  {\bf 1504} (2015) 033, arXiv:1407.0820;\\
%%CITATION = doi:10.1088/1475-7516/2015/04/033;%%
	\textit{P.~Binetruy, J.~Mabillard and M.~Pieroni},
		Universality in generalized models of inflation,
  J. Cosmol. Astropart. Phys. {\bf 1703} (2017) no.03,  060, arXiv:1611.07019
  %doi:10.1088/1475-7516/2017/03/060
 % [arXiv:1611.07019 [gr-qc]].
  %%CITATION = doi:10.1088/1475-7516/2017/03/060;%%

\bibitem{Ruzmaikina}  \textit{T.V.~Ruzmaikina, A.A.~Ruzmaikin},
		Quadratic Corrections to the Lagrangian Density of the Gravitational Field and the Singularity,
 Sov. Phys. JETP \textbf{30} (1970) 372		
		
\bibitem{Gottlober:1993hp}
  \textit{S.~Gottlober, J.P.~Mucket and A.A.~Starobinsky},
  Confrontation of a double inflationary cosmological model with observations,
  Astrophys.\ J.\  {\bf 434} (1994) 417, arXiv:astro-ph/9309049
  %doi:10.1086/174743
   %%CITATION = doi:10.1086/174743;%%



\bibitem{delaCruz-Dombriz:2016bjj}
  \textit{A.~de la Cruz-Dombriz, E.~Elizalde, S.D.~Odintsov and D.~Saez-Gomez},
  Spotting deviations from R$^2$ inflation,
  J. Cosmol. Astropart. Phys. {\bf 1605} (2016) no.05,  060, arXiv:1603.05537
  %doi:10.1088/1475-7516/2016/05/060
   %%CITATION = doi:10.1088/1475-7516/2016/05/060;%%

\bibitem{Wang:2017fuy}
  \textit{Y.C.~Wang and T.~Wang},
  Primordial perturbations generated by Higgs field and $R^2$ operator,
  Phys.\ Rev.\ D {\bf 96} (2017) 123506, arXiv:1701.06636;\\
  %doi:10.1103/PhysRevD.96.123506
   %%CITATION = doi:10.1103/PhysRevD.96.123506;%%
\textit{Y.~Ema},
 Higgs Scalaron Mixed Inflation,
  Phys.\ Lett.\ B {\bf 770} (2017) 403, arXiv:1701.07665;\\
  %doi:10.1016/j.physletb.2017.04.060
  %%CITATION = doi:10.1016/j.physletb.2017.04.060;%%
 \textit{Y.~Ema}, Dynamical Emergence of Scalaron in Higgs Inflation,
  J. Cosmol. Astropart. Phys.  {\bf 1909} (2019) no.09,  027,
  %doi:10.1088/1475-7516/2019/09/027
  arXiv:1907.00993
  %%CITATION = doi:10.1088/1475-7516/2019/09/027;%%
\bibitem{He:2018gyf}
  \textit{M.~He, A.A.~Starobinsky and J.~Yokoyama},
 Inflation in the mixed Higgs-$R^2$ model,
  J. Cosmol. Astropart. Phys. {\bf 1805} (2018) no.05,  064, arXiv:1804.00409
 % doi:10.1088/1475-7516/2018/05/064
   %%CITATION = doi:10.1088/1475-7516/2018/05/064;%%

  \bibitem{Gorbunov:2018llf}
 \textit{D.~Gorbunov and A.~Tokareva},
 Scalaron the healer: removing the strong-coupling in the Higgs- and Higgs-dilaton inflations,
  Phys.\ Lett.\ B {\bf 788} (2019) 37;
 % doi:10.1016/j.physletb.2018.11.015
 arXiv:1807.02392;\\
  %%CITATION = doi:10.1016/j.physletb.2018.11.015;%%
  %%CITATION = ARXIV:1807.02392;%%
\textit{F.~Bezrukov, D.~Gorbunov, C.~Shepherd and A.~Tokareva},
  Some like it hot: $R^2$ heals Higgs inflation, but does not cool it,
  Phys. Lett. B \textbf{795} (2019) 657, arXiv:1904.04737
  %%CITATION = ARXIV:1904.04737;%%
%%%

\bibitem{Karam:2018mft}
  \textit{A.~Karam, T.~Pappas and K.~Tamvakis},
  Nonminimal Coleman-Weinberg Inflation with an $R^2$ term,
  J. Cosmol. Astropart. Phys. {\bf 1902} (2019) 006, arXiv:1810.12884
 % doi:10.1088/1475-7516/2019/02/006
  %%CITATION = doi:10.1088/1475-7516/2019/02/006;%%

\bibitem{Paliathanasis}
  \textit{A.~Paliathanasis, G.~Leon and S.~Pan},
 Exact Solutions in Chiral Cosmology,
  Gen.\ Rel.\ Grav.\  {\bf 51},  no.9 (2019)  106, arXiv:1811.10038;\\
  %doi:10.1007/s10714-019-2594-2
 % [arXiv:1811.10038 [gr-qc]]
  %%CITATION = ARXIV:1811.10038;%%
    \textit{N.~Dimakis, A.~Paliathanasis, P.A.~Terzis and T.~Christodoulakis},
  Cosmological Solutions in Multiscalar Field Theory,
  Eur.\ Phys.\ J.\ C {\bf 79}, no. 7  (2019) 618, arXiv:1904.09713;\\
 % doi:10.1140/epjc/s10052-019-7130-8
  %[arXiv:1904.09713].
   %%CITATION = ARXIV:1904.09713;%%
 \textit{ M.~Zubair, F.~Kousar and S.~Waheed},
Dynamics of scalar potentials in theory of gravity,
  Can.\ J.\ Phys.\  {\bf 97} (2019) no.8,  880.
 % doi:10.1139/cjp-2018-0566
  %%CITATION = doi:10.1139/cjp-2018-0566;%%

\end{thebibliography}
\end{document}